\documentstyle[11pt,amssymb]{article}

\def\be{\begin{equation}}
\def\ee{\end{equation}}

\def\b{\hfil\break}

\title{{\hfill\small\tt Physics\,\,Letters\,{\bf B285},\,217-220\,(1992)}\\
{\hfill}\\
A few comments on N=2 supersymmetric\\
Landau-Ginzburg theories }
\author{A.M. Perelomov\\
{\small\em Institute of Theoretical and Experimental Physics,}\\
{\small\em 117 259 Moscow, USSR}\\
{\small\em and}\\
{\small\em Max-Planck-Institut~f\"ur~Mathematik,}\\
{\small\em Gottfried-Claren-Strasse 26, D-5300 Bonn 3, FRG }
}

\date{}
\begin{document}
\maketitle{}

\begin{abstract}\noindent
In the present note I give two new examples of Landau-Ginzburg theories
considered in detail in recent paper by Cecotti and Vafa.
\end{abstract}

\noindent
In the recent very interesting paper by Cecotti and Vafa [1] they
have considered $N=2$ supersymmetric Landau-Ginzburg theories and have
showed that in many cases the metric for supersymmetric ground states
for special deformations of this metric satisfies the certain system
of PDE's, such as, for example, the Toda equations.

The purpose of the present paper is to give the additional examples
of such theories.\b

\noindent
{\bf 1.} Let us remind first of all some basic facts from $N=2$ supersymmetric
 Landau-Ginzburg theory (for more details, see [1]). The basic quantities
here are the chiral fields $\phi_i$, the vacuum state $|0>$ and the states
\be |j \rangle =\phi_j\,|0\rangle  .\ee
The action of $\phi_j$ on this state is given by the formula
\be \phi_i|j\rangle =\phi_i\phi_j\,|0\rangle =C_{ij}^k\phi_k\,|0\rangle =
C_{ij}^k\,|k\rangle .\ee
So, the action of the chiral field $\phi_i$ in the subsector of vacuum
states is given by the matrix $(C_i)_j^k=C_{ij}^k$. Analogously, we have
anti-chiral fields $\phi_{\bar i}$ and the states $|\bar j\rangle $.
So we may define two metric tensors
\be
\eta_{ij}=\langle j|i\rangle ,\ee
and
\be g_{i\bar j}=\langle \bar j|i\rangle \ee
which should satisfy the condition
\be
\eta^{-1}\,g(\eta^{-1}g)^*=1.\ee
The theory is determined by the superpotential $w(x_a)$ which is
holomorphic function of complex variables $x_a$. The superpotential
completely determines the chiral ring
\be
{\cal R}={\bf C}[x_a]/\partial _a w, \ee
and we may also determine the metric $\eta _{ij}$ by the formula
\be
\eta _{ij}=\langle i|j\rangle =\mbox{Res}_w \,[\phi _i \phi _j], \ee
where
\be
\mbox{Res}_w [\phi ]=\sum _{dw=0} \phi (x)\,H^{-1}(x),\qquad
H=\mbox{det}\,(\partial _i\,\partial _j w). \ee
As for the metric $g_{i\bar j}$, it depends on the parameters $t_1, t_2,
\ldots $ entering to the superpotential $w(x_a)$. As was shown in [1],
it should satisfy the zero-curvature conditions
\be
{\bar \partial }_i\,(g\partial _j\,g^{-1})-[C_j,g(C_i)^+g^{-1}]=0, \qquad
\partial _i=\frac{\partial }{\partial t_i},\quad {\bar \partial }_j
=\frac{\partial }{\partial {\bar t}_j}, \ee
\be
\partial _iC_j-\partial _jC_i+ [g(\partial _ig^{-1}),C_j]- [g(\partial _j
g^{-1}),C_i]=0, \ee
and should satisfy the "reality constraint" (5).

In the paper [1] the many interesting examples of the Landau-Ginzburg
theories were considered. In the next sections we consider two new examples
of such theories.\b

\noindent
{\bf 2.} The model
\be
w(x)= t[\exp (x)- x]. \ee
Here
\be
w'(x)= t[\exp (x)- 1], \ee
and we may identify an element of ${\cal R}$ with the set of the values
of the function $\phi (x)$ at critical points of $w(x)$:
\be
x_j=2\pi ji, \quad j\in {\Bbb Z},\qquad
\phi (x)\in {\cal R} \mapsto \{(\phi )_j\},\quad(\phi )_j=\phi (2\pi ij).\ee
The multiplication operation acts componentwise on ${\phi }$ and we have
also
\be
w''=t\,\exp(x), \ee
and
\be
\mbox{Res}(\phi )=\frac1{t}\,\sum _{j} (\phi )_j. \ee

We choose as basis in $\cal R$ the elements $a_k$ $(k\in{\Bbb Z})$, such that
\be
(a_k)_j=\delta _{kj}. \ee
In this basis we have
\be
\eta _{kl}=\frac1{t}\,\delta _{kl}. \ee
Also,
\be
(C)_l^k=(1-2\pi ik)\,\delta _l^k. \ee
Let us define
\be g_{j\bar k}=\langle {\bar k}|j\rangle . \ee
Then we can see that $w(x)$ is quasi-invariant relative to the translation
operation:
\be T\colon f(x) \to f(x+2\pi i), \ee
\be Tw(x)=w(x)-2\pi i. \ee
So, the metric $g_{i\bar k}$ should be invariant under this transformation,
\be g_{j\bar k}=g_{j+1,\overline {k+1}}, \ee
or
\be g_{j,\overline {j+k}}= g_{0,\bar k}=f_k. \ee
Now, instead the set $\{f_k\}$ we may consider the function
\be f(\theta )=\sum _k f_k\,\exp(2\pi i\theta k), \ee
or
\be g_{k,\bar l}=\int _0^1 f(\theta )\,\exp(-2\pi i(k-l)\theta )\,d\theta .\ee
The reality condition (5) now takes the form
\be |t|^2\,|f(\theta )|^2=1. \ee
Hence,
\be f(\theta)=\frac1{|t|}\,\exp[i\varphi (t,\bar t; \theta)]. \ee
For (9) we have
\be {\bar \partial }\,(f\,\partial f^{-1})-[C,fC^{+}f^{-1}]=0,\qquad
\partial =\frac{\partial }{\partial t},\quad {\bar \partial }=\frac{\partial }
{\partial {\bar t}}. \ee Now \be
C^{+}f^{-1}=[C^{+},f^{-1}]+f^{-1}C^{+}. \ee Hence we have \be
{\bar \partial }(f\,\partial
f^{-1})-[C,f[C^{+},f^{-1}]]-[C,C^{+}]=0, \ee but \be
[C^{+},f^{-1}]=\frac{d}{d\theta }\,(f^{-1}). \ee Hence we have \be
{\bar \partial }\,(f\partial f^{-1})+\frac{d}{d\theta }\,\left(
f\, \frac{d}{d\theta }\,(f^{-1})\right) =0,\ee or \be {\bar
\partial }\,\partial \varphi +\frac{d^2}{d\theta ^2}\,\varphi
=0.\ee Finally \be f=f(t ,{\bar t};\theta
)=\frac{1}{|t|}\,\exp[i\phi (t ,{\bar t};\theta )],\ee \be
\bigtriangleup _3\varphi =0,\ee \be \varphi=\varphi (t,{\bar t};
\theta ),\qquad \varphi (t,{\bar t}; \theta +1)=\phi (t,{\bar t};
\theta ), \ee \be \bigtriangleup _3=\frac{\partial ^2}{\partial
t_1^2}+\frac{\partial ^2} {\partial t_2^2}+\frac{\partial
^2}{\partial \theta ^2},\qquad t=t_1+it_2.\ee Note,that the
variables may be separated in this equation and if you know $\phi
(0,0;\theta )$ or $\phi (t,{\bar t};\theta )$ we may solve this
equation explicitly as $|t|\to \infty $.\b

\noindent
{\bf 3. }The model\b
\be
w= w_c= t\left[ \frac12\,\exp(2x)- 2c\,\exp(x) + x \right],\qquad
c>1. \ee
Here
\begin{eqnarray}
w'(x ) &=& t[\exp(2x) - 2c\,\exp(x) + 1]\nonumber \\
&=& 2t\exp(x)\,(\cosh x- c)\nonumber \\
&=& 2t\,\exp(x)\,(\cosh~x - \cosh~\gamma),\qquad c= \cosh ~\gamma .
\end{eqnarray}
We may identify an element of ${\cal R}$ with the set of its values of the
function $\phi (x)$ at critical points $w(x)$:
\be  \{x_j\}= \{a_j, b_j\},\qquad a_j = -\gamma + 2\pi i j,\qquad
b_j= \gamma+ 2\pi i j, \ee
\be \phi (x ) \in {\cal R} \mapsto \{ (\phi )_j^{a,b} \},\qquad
(\phi)_j^{a,b}= \phi (\mp\gamma + 2\pi i j).\ee

The multiplication operation acts componentwise on $\phi$ and we have also
\be
w''= 2t\,\exp (x)\,[\exp(x)- c]= 2t\,\exp(x)\,[\exp(x)- \cosh \gamma ].\ee
At $x=a_j$ we have $w''=- 2t\,\exp(-\gamma)\, \sinh \gamma $.
\medskip
At $x=b_j$ we have $w''= 2 t\,\exp(\gamma )\,\sinh \gamma $.\b

Here
\be \mbox{Res}(\phi)=\frac1{2t\,\sinh \gamma}\,\sum_j [-\,\exp(\gamma )\,
\phi (a_j)+\exp(-\gamma )\,\phi(b_j)]. \ee

We choose the basis in ${\cal R}$ related to $a_j$ and $b_k$ and in this
basis we have
\be \eta_{k,l}^{a,b}=\mp \,\frac1{2t\,\sinh \gamma}\,\exp(\pm \gamma)\,
\delta_{k,l}.\ee
Also
\be (w_{j})^{a,b}= t\,(A \pm B + 2\pi ij),\ee
where
\be  A= -\,(1+\frac12\,\cosh 2\gamma),\qquad B=\frac12\,\sinh 2\gamma -
\gamma .\ee
Hence,
\be C=\left( \begin{array}{cc}C^a&0\\ 0&C^b \end{array}\right) .\ee
\be
(C^a)_l^k=(A+ B+ 2\pi i k)\,\delta_k^l,\quad
(C^b)_l^k=(A- B+ 2\pi i k)\,\delta_k^l.\ee

The matrix $g$ has now the block form \be g=\left(
\begin{array}{cc} g^{aa}&g^{ab}\\ g^{ba}&g^{bb} \end{array}
\right) , \qquad g^{aa}=\{g_{j,k}^{a,a}\}, \ldots . \ee The
invariance group in this case is generated by the translation \be
T\colon x\to x+ 2\pi i.\ee Hence \be
g_{j+l,\bar{k}+\bar{l}}^{a,a}= g_{j,\bar{k}}^{a,a},\ldots ,\ee and
\be g^{aa}(\theta)=\sum \exp(2\pi i(k- j)\,\theta )\,g_{j,\bar
{k}},\ldots ,
\qquad g(\theta)=\left( \begin{array}{cc} g^{aa}(\theta )& g^{ab}(\theta )\\
g^{ba}(\theta ) & g^{bb}(\theta )\end{array} \right) \,. \ee

In these notations,
\be \eta=\frac1{2t\,\sinh \gamma}\,\left( \begin{array}{cc} -\exp(\gamma )&0\\
0& \exp(-\gamma )\end{array}\right) ,\ee
\be C= \left( AI+ B \Sigma_3 + \frac{d}{d\theta}\right) ,\qquad
C^+=\left( AI + B\Sigma_3-\frac{d}{d\theta}\right) .\ee
The reality condition should be taken in the form
\be \left( \eta^{-1}\,g(\theta)\right) \left( \eta ^{-1}\,g(-\theta)\right)
^*= I.\ee

It is easy to show that this condition is equivalent to \be \left(
\eta_0^{-1}\,\tilde g(\theta )\right) \left( \eta_0^{-1}\, \tilde
g(-\theta )\right) ^*=I,\qquad \eta_0=\frac1{2t\,\sinh\gamma}\,
(-\Sigma_3).\ee Here \be g=D\tilde g D,\qquad D=\left(
\begin{array}{cc} \exp\left( -\,\frac12\, \gamma \right) &0\\
0&\exp \left( \frac12\,\gamma \right) \end{array} \right) . \ee

So, up to a normalization factor, we may consider that $\tilde g\in SU(1,1)$.
The equation (3) may be reduced now to the equation for the matrix
$\tilde g$:
\be {\bar\partial }\,(\tilde g\,\partial {\tilde g}^{-1})-[C, \tilde g
[C^+,\tilde g^{-1}]]= 0,\qquad C=\frac{d}{d\theta}+B \Sigma_3,\quad
C^+=-\frac{d}{d\theta}+B\Sigma_3,\ee
or
\be
{\bar\partial }\,(\tilde g\,\partial\tilde g^{-1})+\left[ \frac{d}{d\theta}
+B\Sigma_3,\,\tilde g\left[ \frac{d}{d\theta}- B\Sigma_3, \tilde g^{-1}
\right ] \right ]=0.\ee

Let
\be \tilde g\,\partial\tilde g^{-1}= A_t,\qquad \tilde g\,\frac{d}{ d\theta}\,
\tilde g^{-1}=A_{\theta}.\ee

We have also
\be \frac{d}{d\theta}\,\tilde g^{-1}= \tilde g^{-1}\, A_{\theta},\qquad
\frac{d}{d\theta}\,\tilde g=-\,A_{\theta}\, \tilde g.\ee

Finally we have
\be \bar\partial A_{t}+ \frac{d}{d\theta}\,A_{\theta}- B\left[ \left(
\tilde g\,\Sigma_3\,\tilde g^{-1}-\Sigma_3\right) ,A_\theta\right] -
B^2\left[ \Sigma_3,\tilde g\Sigma_3 \tilde g^{-1}\right]=0, \ee
where
\begin{eqnarray}
&& A_{t}=\tilde g\,\partial\tilde g^{-1},\qquad
A_{\theta}=\tilde g\,\frac{d}{d\theta}\,\tilde g^{-1},\qquad B=\frac12\,
\sinh{2\gamma}-\gamma,\nonumber \\
&& {\tilde g}\in SU(1, 1),\qquad A_{t},\,A_{\theta }\in su(1,1),\nonumber \\
\tilde g\,\Sigma_3\,\tilde g^{-1}\in su(1,1).\end{eqnarray}

Note that for $B \to 0$ (this corresponds to the case of one chain
with double zeros) we obtain the equation for the  principal
chiral field in three dimensions with coordinates $t_1$, $t_2$ and
$\theta $ for the group $SU(1, 1)$ (see [1]): \be
\partial_{\mu}\tilde g\,\partial_{\mu}\tilde g^{-1}= 0,\qquad
\mu=1,2,3.\ee Here we considered the case of two chains of zeros.
The consideration of an arbitrary finite number of chains gives an
analogous equation for some real simple Lie algebra.

This work was finished during the author's stay at the Max-Planck-Institut
f\"ur Mathematik. I would like to thank Professor F.Hirzebruch for his kind
hospitality and Prof. C.Vafa for bringing problems of Landau-Ginzburg
theories to my attention.

\end{document}